\documentclass[%
reprint,
superscriptaddress,
showpacs,preprintnumbers,
nofootinbib,
 amsmath,amssymb,
 aps,
prb,
]{revtex4-1}

\usepackage{graphicx}
\usepackage{dcolumn}
\usepackage{bm}

\begin{document}

\title{Pressure effect on the energy structure and superexchange interaction of the undoped orthorhombic La$_2$CuO$_4$: beyond the low-energy approximation}
\author{Vladimir A. Gavrichkov}
\affiliation{L. V. Kirensky Institute of Physics, Siberian Branch of Russian Academy of Sciences, 660036, Krasnoyarsk, Russia}
\affiliation{Siberian Federal University, 660041, Krasnoyarsk, Russia}

\author{Zlata V. Pchelkina}
\affiliation{Institute of Metal Physics, Ural Branch of the Russian Academy of Sciences, 620219, Ekaterinburg, Russia}
\affiliation{Theoretical Physics and Applied Mathematics Department, Ural Federal
University, 620002 Ekaterinburg, Russia}

\author{Igor A. Nekrasov}
\affiliation{Institute of Electrophysics, Ural Branch, Russian Academy of Sciences,  620016, Ekaterinburg, Russia}

\author{Sergey G. Ovchinnikov}
\affiliation{L. V. Kirensky Institute of Physics, Siberian Branch of Russian Academy of Sciences, 660036, Krasnoyarsk, Russia}
\affiliation{Siberian Federal University, 660041, Krasnoyarsk, Russia}
\date{\today}

\begin{abstract}

Using LDA+GTB multi-band approach, we studied the compression dependence of the electronic structure and in-plane superexchange interaction
$J(P)$  in the antiferromagnetic La214 at the 0\% and 3\% - hydrostatic and unaxial (along c axial) compression. We obtained the superexchange interaction $J(P=0)\thickapprox0.15eV$ is enhanced by $\thicksim$ 20\% under the 3\% - hydrostatic compression and vice versa the $J(P)$ is decreased slightly by $\thicksim$ 5,7\% under the 3\% - uniaxial compression. In both cases the $J(P)$ correlates with the in-plane hopping parameters and $dd$-excitation energy $ \delta_s=\varepsilon(^3B_{1})-\varepsilon(A_{1})$ involving the the two-hole states: Zhang-Rice  singlet and ${}^3{B_{1}}$ triplet states. The spectral density of the first removal states is a combined singlet-triplet character and a sign of changes in the one with the pressure clearly reproduces the $\vec{k}$-distribution of quasiparticle states with a different $a_1$- and $b_1$-symmetry over the Brillouin zone as a whole.


\end{abstract}


\maketitle

\section{\label{sec:intr}Introduction\\}
A superexchange study in the high $T_c$ cuprates is an important part in the bosonic battle: ~\cite{Lanzara_etal2001, Hwang_etal2004, Lee_etal2006, Carbotte_etal1999, Demler_etal1998, Abanov_etal2002, Scalapino_2012, Cuk_etal2004} magnetic or lattice - what are relevant to pairing? The studies of the different pressure dependences of critical temperature $T_C(P)$ observed universal trend: ${{\partial {T_C}} \mathord{\left/
{\vphantom {{\partial {T_c}} {{P_c}}}} \right.
\kern-\nulldelimiterspace} {{P_c}}} < 0$ and ${{\partial {T_C}} \mathord{\left/
{\vphantom {{\partial {T_c}} P}} \right.
\kern-\nulldelimiterspace} P} > 0$ ~\cite{Hardy_etal2010} for an anisotropic and isotropic pressures respectively.
A correlation between the Cu-O$_{ap}$ apical bond distance and $T_c$ in the cuprates has been found by Jorgensen and co-workers. \cite{Jorgensen_etal1996} The Cu-O$_{ap}$ apical
bond distance increases with the $c$-axis length. The role of the apical oxygen on the electronic properties has
also been addressed in several theoretical works ~\cite{Sakakibara_etal2010, Ohta_etal1991, Feiner_etalPRL1996, Pavarini_etal2001} which
all find that $T_c$ is highest for materials in which the
apical bond distance is large. The Y-123, where
contrary to the general trend for cuprates, $T_c$
increased against the contraction along all the crystal axes, is the exception. Due to a presence of CuO chains its behavior is unusual, and not representative of the cuprates  as a whole.\cite{Hardy_etal2010}

Recently, the time-frequency resolved spectroscopy~\cite{Dal_Conte_etal2012} has shown the dominant role of the non-redarding electronic mechanism of pairing in optimally doped Bi-cuprate. The exchange magnetic interaction is one of the candidates for electronic pairing. At the same time the experimental studies ~\cite{Aronson_etal1990, Eremets_etall1991, Schilling2007} show increasing superexchange ${{\partial J} \mathord{\left/
{\vphantom {{\partial J} {\partial P > 0}}} \right.
\kern-\nulldelimiterspace} {\partial P > 0}}$ under hydrostatic pressure $P$. However, we didn't find any publication on the $J$ dependence on uniaxial pressure.

Our theoretical work will be devoted the issue that could not be solved at low-energy limit: the superexchange interaction and as an example its different  pressure dependences $J(P)$ in the undoped La214. This phenomena cannot be understood relying only on the low-energy approximation (in the framework of the three-orbital $pd$-model).\cite{Varma_etal1987, Emery_1987} The value of superexchange interaction $J$ in the La214 is controlled by a large number of excited two hole states: $N_T$ triplets and $N_S$ singlets including the Zhang-Rice state $A_{1}$.

Essentially, there are two acceptable approaches to the study of superexchange interaction. The first is the calculation with the intermediate two-hole states which arise through hopping from oxygen to oxygen in the perturbation theory of a higher order than a fourth.\cite{Eskes_etal1993} Another approach is a cell perturbation theory taking into account all of the excited states. The latter seems more appropriate,\cite{Jefferson_etal1992,Eskes_etal1993, Feiner_etal1996} where, however, it is necessary to work with a large number of excited states.\cite{Gavrichkov_etal2008}  Especially, if we keep in mind that the energy gap in the La214 between the $A_{1}$ singlet and $^3B_{1}$ triplet two-hole cell states involving Cu-$d_{z^2}$ and O$_{ap}$-$p_z$ apical oxygen states can be quite small. \cite{ Kamimura1987, Kamimura_etal1990, Eskes_etal1991, Grant_etal1991, Ohta1_etal1991}

Using  the LDA+GTB approach ~\cite{Ovchinnikov_etal2012} which extends the cell perturbation theory ~\cite{Jefferson_etal1992, Feiner_etal1996} to an arbitrary number of the excited cell states, we calculate the compressure dependence of superexchange constant $J(P)$ in the orthorhombic La214, where unlike the pressure dependencies of superexchange interaction  in many other conventional transition-metal oxide,\cite{JohnsonSievers1974, Kaneko_etal1987, KimMoret1988, Massey_etal1990}
the two-magnon Raman scattering experiments show that $J(P)$ has a substantially weaker pressure dependence.\cite{Aronson_etal1990, Eremets_etall1991, Schilling2007}
Within the perturbation theory using the atomic orbitals representation,\cite{ZhangRice1988, Eskes_etal1993, Maekawa_etal2004} a superexchange interaction is obtained at the fourth-order of a perturbation theory and the weak-pressure dependence of $J(P)$ is clearly not consistent with the pressure dependencies of the main parameters of the $pd$-model:
$t_{pd}\sim a^{-\alpha}$ $(2.5 \lesssim \alpha \lesssim 3.0)$, \cite{Smith_1969, Fuchikami_1970, Jough_1975, Shrivastava_etal1976}
$\Delta\sim a^{-\beta}$ $(\beta\approx0.4\pm0.4)$.\cite{Venkateswaran_etal1989} Comparison of the results
at the the fourth-order with the calculations in higher orders of perturbation theory ~\cite{Eskes_etal1993} and the exact diagonalization of finite clusters ~\cite{ Eskes_etal1989, Ohta_etal1991, Stechel_etal1988, Annet_etal1989, Eskes_etal1993} shows that the in-plane superexchange $J$ depends on the $t_{pd}$ significantly weaker and, because of the too large value of  $t_{pd}/\Delta$ in the CuO$_2$ layer, the fourth order may be insufficient.

We discuss the exchange constant $J$, and compare our results with the conclusion from the neutron experiments in undeformed La214,\cite{Coldea_etal1990} the experiments related to the two-magnon Raman scattering in deformed materials ~\cite{Aronson_etal1990} at the 0\%, 3\% hydrostatic and uniaxial (along c axis) compressions.

One of the features of the study is that the exchange interaction is derived from the original electronic structure of material, and we can compare a compression effect on the superexchange interaction with a same effect on the electronic structure.
All related changes in the electronic structure of a material under pressure is also available to study.

At a fixed energy of the incident photons a photocurrent $I$ in the unpolarized ARPES experiments is proportional to the partial contribution of the spectral density from electron orbital $\lambda$:
\begin{equation}
I\left( {\vec k,E,P} \right) \sim { {I_0\left(\vec{k},P \right)}}f_F(E)\sum\limits_\lambda  {{A_{_\lambda }}\left( {\vec k,E,P} \right)},
\label{eq:1}
\end{equation}
where $A_{_\lambda }\left( \vec{ k},E,P \right)$, $f_F(E)$, $I_0\left(\vec{k},P \right)$ are the partial spectral density from the $\lambda$-orbital, the Fermi function and the matrix element of the interaction of electron with the incident photons respectively. Over a pressure range where a crystal symmetry is unchanged, the matrix element $I_0\left(\vec{k},P \right)\approx I_0\left(\vec{k} \right)$, i.e. the relative changes of the photocurrent $\delta I(\vec{k},E,P)$  with a pressure are caused by the changes in the spectral intensity $\delta {A_{tot}}\left( {\vec k,E,P} \right) = {A_{_{tot}}}\left(\vec{k},E,P \right) - {A_{_{tot}}}\left(\vec{k},E\right)$, where ${A_{tot}}\left( {\vec k,E,P} \right) = \sum\limits_\lambda  {{A_{_\lambda }}\left( {\vec k,E,P} \right)}$.

In this work, we study the compressure effects on the ${A_{tot}}\left( {\vec k,E,P} \right)$ along with changes in the superexchange interaction $J(P)$  in a framework of the five orbital model
where the orbital index $\lambda$ corresponds either to the Cu$3d$ orbitals
$d_{x^2-y^2}$ and $d_{3z^2-r^2}$ or to symmetrized combinations of the O-$2p^\sigma$ atomic orbitals centered at the copper site $\vec{R}_f$: $b, a$ orbitals (transforming like $b_1$ and $a_1$ ), \cite{Feiner_etal1996} O$_{ap}$-$2p_z$.
To specify partial contribution ${A_{{b_1}}} = \sum\limits_{\lambda} {{A_{_\lambda }}\left( {P} \right)}$ and ${A_{{a_1}}} = \sum\limits_{\lambda'} {{A_{\lambda'}}\left( {P} \right)}$ to the $A_{tot}(\vec{k},E,P)$-total spectral density of the first removal electron state ($frs$) in the undoped antiferromagnetic La214, we also studied compression dependence of the partial contributions. In line with our results
the spectral density from $b_1$- and $a_1$-symmetry quasiparticle ($qp$ -) states extends along the edges of the AFM Brillouin zone and the near the $\vec{k}$-points: $(0,0)$, $(\pi,\pi)$ respectively. The sign of the pressure effects on the total spectral density $\delta A_{tot}\left( {\vec k,E,P} \right) $ provides a clear imprints of the $b_1$- and $a_1$- contributions to the $frs$-state over the Brillouin zone as a whole.

We obtained that the superexchange constant in the undeformed La214 is close to the experimental value 0.146 $eV$ ~\cite{Coldea_etal1990} and increases by $\thicksim20\%$ under 3\%-hydrostatic compression. At the same time, the superexchange interaction is only slightly reduced by $\thicksim-5.7\%$ under the uniaxial compression. According to the available experimental results,\cite{Aronson_etal1990} the superexchange interaction is increased by $\thicksim18\%$ at 3\%-hydrostatic compression ($P\thicksim205$Kbar). In both cases, the hydrostatic and anisotropic compression the $J(P)$ correlates with the energy $\delta_s=\varepsilon(^3B_{1})-\varepsilon (A_{1})$ of $dd$-excitation involving the the two-hole states:  Zhang-Rice state $A_1$ and triplet state ${}^3{B_{1}}$. Showing with the $T_c(P)$ the similar trend:  ${\raise0.7ex\hbox{${\partial J}$} \!\mathord{\left/
 {\vphantom {{\partial J} {\partial P}}}\right.\kern-\nulldelimiterspace}
\!\lower0.7ex\hbox{${\partial P}$}}>0 $ and ${\raise0.7ex\hbox{${\partial J}$} \!\mathord{\left/
 {\vphantom {{\partial J} {\partial P}}}\right.\kern-\nulldelimiterspace}
\!\lower0.7ex\hbox{${\partial P_c}$}}<0 $ the $J(P)$ dependence support the magnetic view of a discussion on pairing mechanism at least in the single layer cuprates.

We also carried out the GTB calculation of a hypothetical case of the $A_{1} \leftrightarrow ^3B_{1}$ two-hole state crossover.
Due to the orbital features of Zhang-Rice state the superexchange keeps the antiferromagnetic character even at such a hypothetical set of the parameters of Hamiltonian.

\section{\label{sec:II} All valence states in multiband $pd$-model\\}

In the multiband $pd$-model,\cite{Gaididei_etal1988} a Hamiltonian includes the local energies of holes for the oxygen and copper at the different orbital states, the intraatomic Coulomb and exchange interactions for copper and oxygen, hoppings, and the copper-oxygen Coulomb interaction. The important difference with the low energy three orbital $pd$-model \cite{Varma_etal1987, Emery_1987} is related to an addition of the $z$-oriented $d_{z^2}$ orbital of copper and $p_z$ orbital of the apical oxigen ions. In the framework of the local density approximation in combination with the generalized tight-binding method (LDA+GTB), the Hamiltonian parameters are calculated from first principles.\cite{Korshunov_etal2005} Then, the cell approach of the generalized tight-binding method ~\cite{Gavrichkov_etal1998, Ovchinnikov_etal2012} is used to take into account strong electron correlations explicity. A crystal lattice is divided into unit cells, so that the Hamiltonian is represented by $H_0+H_1$, where the component $H_0$ is the sum of intracell terms and the component $H_1$ takes into account the intercell hoppings and interactions. The component $H_0$ is exactly diagonalized. The exact multielectron cell states $|n,\nu\rangle$ and energies $\xi_{n\theta}$ are determined. Then these states are used to construct the Hubbard operators of the unit cell $\vec{R}_{f}: X^{n\theta,n'\theta'}_f = |n\theta\rangle\langle{n'}\theta'|$, where $\theta=S(0), M(-1, 0, +1)$ for the singlet and triplet states respectively, the index $n$ is the sequence number of doublet state in the one-hole sector $n_h=1$ and also singlet, triplet states in the two-hole sector $n_h=2$ (Fig.\ref{fig:1}). Thereafter, the component $H_1$ is exactly written in the $X$ - operator representation and the intercell interactions are included in terms of the perturbation theory.
The procedure and results of calculations for the undeformed
CuO$_2$ layer are described in our previous review paper.\cite{Ovchinnikov_etal2012}
In the X-operator representation the component $H_0$ is determined by the sum over the unit cells, that is
\begin{widetext}
\begin{equation}
H_0=\sum_{f}
\left\{%
\varepsilon_0X^{00}_f+\sum_{l\sigma}\left(\epsilon_l-\mu\right)X_f^{l\sigma,l\sigma}
+\left[\sum_{n=1}^{N_S}(E_{nS}-2\mu)X_f^{nS,nS}+\sum_{m=1}^{N_T}\sum_{M}(E_{mM}-2\mu)X_f^{mM,mM}\right]
\right\},
\label{eq:2}
\end{equation}
\end{widetext}

\begin{figure}
\includegraphics{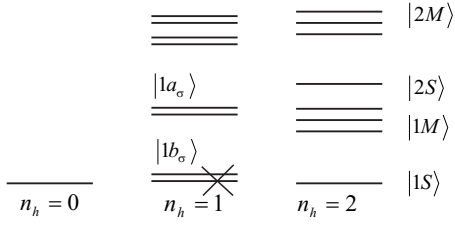}%
\caption{\label{fig:1} Energy-level scheme of the Hilbert space of the CuO$_6$ unit cell with the numbers of holes $n_h =$
0, 1, and 2. The cross indicates the occupied ground state $|1b_{\sigma}\rangle$
of the configuration $d^{9}p^{6} + d^{10}p^{5}$ in the undoped case.}
\label{fig:1}
\end{figure}

where $\varepsilon_0$ is the energy of the "vacuum" term $d^{10}p^6$ in the hole representation, $\epsilon_l$ is the energy of one-hole molecular orbitals with the spin projections $\sigma=\pm1/2$, and the index $l$ runs over all one-particle states of the CuO$_6$ cluster. The term in square brackets in (\ref{eq:2}) describes the contributions from two-hole singlet states $|n,S\rangle$ and the triplet states $|m, M\rangle$. The indices $n (1\leq n \leq N_S)$ and $m (1 \leq m \leq N_T)$ numerate all two-hole spin singlets and triplets respectively.
A completeness of the set of local Hubbard operators is represented by the sum:
\begin{equation}
X_f^{00}+\sum_{l\sigma } X_f^{l\sigma,l\sigma} + \sum_n {X_f^{nS,nS}} + \sum_m\sum_M X_f^{mM,mM} =1.
\label{eq:3}
\end{equation}
The energy-level scheme of the many-electron terms of the $H_0$ for the La$214$ compound is depicted in Fig.\ref{fig:1}. A creation of an electron at the bottom of the conduction band is determined by the matrix element $\gamma_{\lambda\sigma}\left(0\right) = \left\langle0\right|a_{f\lambda\sigma}\left|b_{\sigma}\right\rangle$. A creation of the hole upon $p$-type doping is determined by the matrix elements with the participation of all two-hole terms:
\begin{equation}
\gamma_{\lambda\sigma}(n)=\langle b_{\bar{\sigma}}|a_{f\lambda\sigma}|ns\rangle,
\nonumber
\end{equation}
\begin{equation}
\gamma_{\lambda\sigma}(m)=\langle b_{\sigma}\left|a_{f\lambda\sigma}\right|{m,+2\sigma} \rangle.
\label{eq:4}
\end{equation}
In the X-operator representation, the hole annihilation operator has the form
\begin{eqnarray}
a_{f\lambda \sigma } = \gamma _{\lambda \sigma }(0)X_f^{0,\sigma } + \sum_ n \gamma _{\lambda \sigma }(n)X_f^{ \bar{\sigma} ,\;ns}\nonumber\\
+ \sum_m \gamma _{\lambda \sigma }(m) \left( X_f^{\sigma ,\;m2\sigma } + \frac{1}{\sqrt 2 }X_f^{ \bar{\sigma},\;m0} \right).
\end{eqnarray}

Forming the singlet
and triplet Hubbard subbands with a hybridization between them the two-hole states with nonzero matrix elements (\ref{eq:4}) are involved to a formation of the valence band energy structure. It should be noted that the singlet (triplet) band is a conventional name for designation of the electronic band with the spin $\sigma = 1/2$ but with a participation of final singlet (triplet) terms. A representation of the off-diagonal operators $X$ can be simplified by introducing the root vectors $\vec \alpha _r(n\theta,{n'}\theta')$ corresponding to a pair of the initial and final states. In this notation, the last relationship takes the form
$a_{f\lambda \sigma } = \sum_r \gamma _{\lambda \sigma }(r) X_f^r$
where the integer index $r$ numbers all one-particle excitations:

\begin{equation}
\{\vec{\alpha}_r\}  = \{ (0,l{\sigma} );(l{\sigma} ,ns);(l\sigma,m2\sigma' );( l{\sigma},m0)\}.
\label{eq:6}
\end{equation} \\
Moreover, in the same notation, the Hamiltonian of the intercluster hopping has the simple form
\begin{eqnarray}
{H_1} &&= \sum\limits_{fg} {} \sum\limits_{\lambda \lambda '\sigma } {} t_{fg}^{\lambda \lambda '}a_{g\lambda \sigma }^ + {a_{g\lambda '\sigma }} + h.c. \nonumber \\
&&= \sum\limits_{fg} {} \sum\limits_{rr'} {} t_{fg}^{rr'}\mathop {X_f^r}\limits^ +  X_g^{r'}
\label{eq:7},
\end{eqnarray}
where $t_{fg}^{\lambda \lambda'}$ is the matrix of hopping integrals of a hole from the $g$-th cell (in the orbital state $\lambda'$) to the $f$-th cell (in the orbital state $\lambda$) and
\begin{eqnarray}
t_{fg}^{rr'}& = &
 \sum\limits_{\lambda \lambda '} \sum\limits_\sigma  t_{fg}^{\lambda \lambda '}\nonumber\\
 &\times &\left[ \gamma _{\lambda \sigma }^*\left( r \right)\gamma _{\lambda '\sigma }\left( r' \right) + \gamma _{\lambda '\sigma }^*\left( r \right)\gamma _{\lambda \sigma }\left( r' \right) \right].
 \label{eq:8}
\end{eqnarray}

Since each index $r$ characterizes the band of quasiparticles in a strongly correlated system (the Hubbard band index), the diagonal terms $t^{rr}$ in the last expression describe the dispersion of the $r$-th band and the off-diagonal terms $t^{rr'}$ describe the hybridization of the
$r$-th and $r'$-th bands.
The equations of motion for the Green's function
can be solved within the different approximations.
In the diagram technique for $X$-operators~\cite{Ovchinnikov_etal2004} with the intercell hopping $H_1$ as a pertubation the Hartree-Fock approximation results in the Hubbard-I type solution

\begin{widetext}
\begin{equation}
D_{\vec k}^{rr'} = {\delta _{rr'}}{\delta _{fg}}D_r^{\left( 0 \right)} + D_r^{\left( 0 \right)}\sum\limits_{\lambda \lambda ' \sigma} {\sum\limits_{r''} {{t_{\lambda \lambda '}}\left( {\vec k} \right)\left\{ {\gamma _{\lambda \sigma}^*\left( r \right){\gamma _{\lambda ' \sigma}}\left( {r''} \right) + \gamma _{\lambda ' \sigma}^*\left( r \right){\gamma _{\lambda \sigma}}\left( {r''} \right)} \right\}D_{\vec k}^{r''r'}} },
\label{eq:9}
\end{equation}
\end{widetext}
where  ${t_{\lambda \lambda '}}\left( {\vec k} \right) = \sum\limits_{\vec h} {{t_{\lambda \lambda '}}\left( {\vec h} \right){e^{i\vec k\vec h}}} $.
We can use the matrix notation
\begin{equation}
{\hat D_{\vec k}} = {\hat \Pi ^{ - 1}}\left( {\vec k} \right){\hat D^{\left( 0 \right)}},
\label{eq:10}
\end{equation}
where  $\hat \Pi \left( {\vec k} \right) = 1 - {\hat D^{\left( 0 \right)}}{\hat t}\left( {\vec k} \right)$ and
\begin{equation}
D_{0fg}^{rr'} = {\delta _{fg}}{\delta _{rr'}}\frac{{F\left( r \right)}}{{E - {\Omega _r}}} ,
\label{eq:11}
\end{equation}
$F\left( r \right) = \left\langle{ X_f^{n\theta,n\theta} + X_f^{n'\theta',n'\theta'}} \right\rangle$ and ${\Omega _r} = \Omega \left( {{{\vec \alpha }_r}} \right) = {\xi _{n\theta }} - {\xi _{n'\theta '}}$.
Thus the dispersion relations of the quasiparticles are determined by an equation on the poles of matrix Green function ${\hat D_{\vec k}}$ :
\begin{equation}
\left\| {\left( {E - {\Omega _r}} \right){\delta _{rr'}} - F\left( r \right)t^{rr'}\left( {\vec k} \right)} \right\| = 0.
\label{eq:12}
\end{equation}
Each $r-th$ root vector defines the Fermi excitation in multielectron system of the CuO$_2$ layer - quasiparticle with charge $e$, spin 1/2, and local energy $\Omega _r$.
Table \ref{tab:1} shows the values of hopping parameters and single electron energies for orthorhombic La214 obtained in the frameworks of Wannier function projection procedure for different sets of
trial orbitals ~\cite{Korshunov_etal2005} at zero, 3\% - hydrostatic and uniaxial compressions.
Despite the fact that the table shows the same vector, a system of the connecting vectors
varies slightly with increasing compression, and a volume of the unit cell under uniaxial compressure was assumed a constant.

\begin{figure*}
\includegraphics{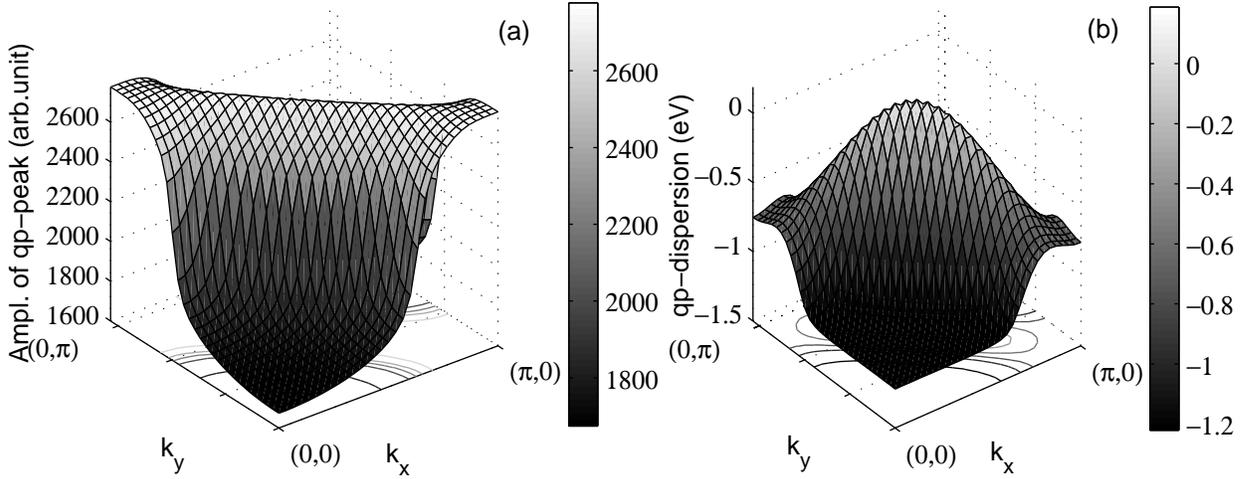}%
\caption{\label{fig:2} (a): $\vec{k}$ - dependencies of the $qp$-peak amplitude $A_{tot}(\vec{k},E_{0\vec{k}\sigma})$  and (b): $qp$-peak position $E_{0\vec{k}\sigma}$ of the first removal electron states in the undeformed La214}
\end{figure*}

\begin{figure*}
\includegraphics{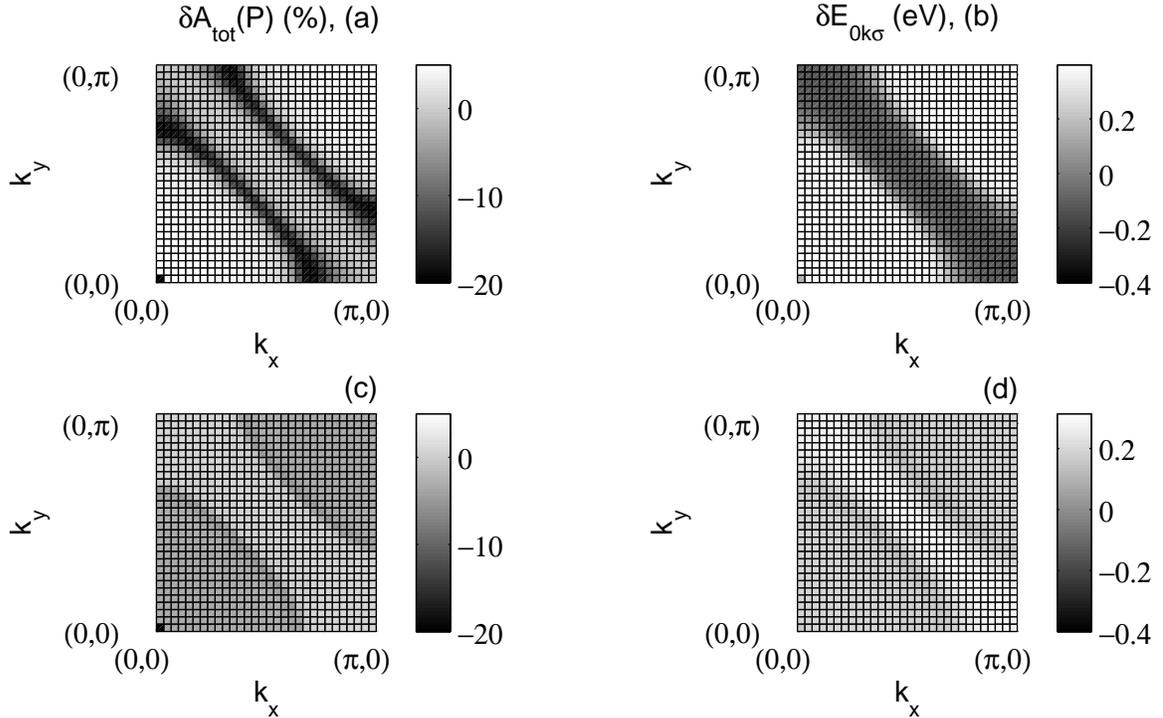}%
\caption{\label{fig:3} (a) and (c): $\vec{k}$ - dependence of the $\delta A_{tot}(P)$-relative changes in a spectral intensity at the $3\%$ - hydrostatic and uniaxial compressions respectively, (b) and (d): $\vec{k}$ - dependence of the $\delta E_{0 \vec{k} \sigma}$-changes in the $qp$-peak position at the $3\%$ - hydrostatic and uniaxial compressions respectively.}
\end{figure*}

In the Russell-Saunders scheme all the possible solutions $E_{rk\sigma}$ are classified according to spin $\sigma$ of the quasiparticle and the number of solutions is equal to twice the number of root vectors $r$: $2N$, where $N=N_S+3N_T$. Because of a spin degeneracy of the ground state of the AFM in the single-hole sector, there is a symmetry in the $E_{rk\sigma}$ with respect to the replacement $\sigma\leftrightarrow\overline{\sigma}$. The $E_{r=0k\sigma}$-energy position of the $qp$-peak of the $frs$-state corresponds to the solution with the lowest energy  in a hole representation. A spectral density of the $qp$-states (amplitude of the $qp$-peak) in turn is determined by the single-particle Green's function
 \begin{equation}
 {A_{tot} }\left( {\vec k,E,P} \right) = \left( { - \frac{1}{\pi }} \right)\sum\limits_{\lambda, \sigma}  {{\gamma _{\lambda \sigma }}\left( r \right)\gamma _{\lambda \sigma }^*\left( {r'} \right){\mathop{\rm Im}\nolimits} D_{\vec k}^{rr'}}.
 \label{eq:13}
\end{equation}
According to the equations (\ref{eq:12}) and (\ref{eq:13}) the $A_{tot}(\vec{k},E=E_{0\vec{k}\sigma})$ - amplitude  and $E_{0k\sigma}$ - energy position of  a $qp$-peak for the $frs$-state  in the undeformed AFM La214 behaves as follows (Fig.\ref{fig:2}). The $frs$-state has a mixed singlet-triplet character with the $\vec{k}$-depending amplitude of a $qp$-peak. The latter has a maximum value along the antiferromagnetic Brillouin zone edges, because the number of $qp$-states in the initial singlet and triplet bands differs significantly. Under  hydrostatic compressure the $frs$-band width increases and amplitude of a $qp$-peak near the $\vec{k}$-points $(0,0)$ and $(\pi,\pi)$ significantly attenuated (see $ \delta A_{tot}(P)$ and $\delta E_{0\vec{k}\sigma}(P)=E_{0\vec{k}\sigma}(P)-E_{0\vec{k}\sigma}(0)$ on Fig. \ref{fig:3},(a) and (b)). Under uniaxial compressure the $frs$-band width decreases and amplitude of a $qp$-peak near the $\vec{k}$-points $(0,0)$ and $(\pi,\pi)$ is increased. As consequence, the total spectral density over the Brillouin zone is leveled (Fig. \ref{fig:3}, (c) and (d)). A significant contribution of the $a_1$-orbital group at the $(0,0)$ and $(\pi,\pi)$ $\vec{k}$-points of Brillouin zone is a reason of this spectral intensity behavior.

\section{\label{sec:III}Effective superexchange hamiltonian\\}
\begin{figure}
\includegraphics{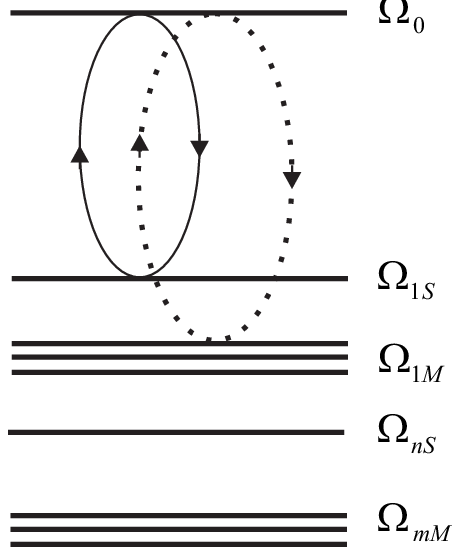}%
\caption{\label{fig:4} The virtual excitations from the occupied valence band states into the empty conductivity band and back that results in the superexchange interaction in La214. Solid line corresponds to the antiferromagnetic contribution forming by the singlet bands, and dotted line denotes the ferromagnetic contribution of triplet bands}
\end{figure}

The superexchange interaction appears at the second
order of the cell perturbation theory with respect to hoppings.\cite{Jefferson_etal1992}
That corresponds to virtual excitations from the
occupied singlet and triplet bands through the insulating
gap to the conduction band at the root vector $r = 0,\;\;{\alpha _0} = (0,\sigma )$ and back (Fig.\ref{fig:4}).
These perturbations are described by the off-diagonal
elements $t_{fg}^{0r}$ with $r \ge 1$ in expression (\ref{eq:8}). In the Hubbard
model, there is only one such element $t^{01}$, which
describes the hoppings between the lower and upper
Hubbard bands. In our case, the set of nonzero matrix
elements $\gamma _{\lambda \sigma }(r)$ with $r \ge 1$ determines the interband hoppings.
In order to eliminate them, we generalize the
projection operator method proposed by Chao et al ~\cite{Chao_etal1977}
to the Hubbard model. Since the diagonal Hubbard
operators are projection operators, the $X$-operator representation
allows us to construct this generalization. In
our case the total number of diagonal two-hole operators $X_f^{\mu\mu}$
is equal to $N$ and the  sequence index $\mu$ ($1 \le \mu  \le N$) runs over all the two-hole states.

By disregarding the exponentially
low temperature occupation of excited one-hole terms
in the absence of doping when none of the two-hole
state is occupied, we can retain only one lower one-hole
 state $\left| {{1b_\sigma }} \right\rangle$ marked by the cross in Fig.\ref{fig:1}. Further we will omit the index $1b$ in the set of root vectors (\ref{eq:6}).

We choose a pair of neighboring cells $(i, j)$ and construct
the set of projection operators ${p_\mu }$:
\begin{equation}
{p_0} = \left( {X_i^{00} + \sum\limits_\sigma  {X_i^{\sigma \sigma }} } \right)\left( {X_j^{00} + \sum\limits_\sigma  {X_j^{\sigma \sigma }} } \right),
\label{eq:14}
\end{equation}
\begin{equation}
{p_\mu } = X_i^{\mu \mu } + X_j^{\mu \mu } - X_i^{\mu \mu }\sum\limits_{\nu} {X_j^{\nu\nu}}.
\label{eq:15}
\end{equation}
It is easy to check that each operator ${p_\mu }$ is a projection
operator $p_\mu ^2 = {p_\mu }$, and these operators form a complete system and are
orthogonal, ${p_\mu }{p_\nu } = {\delta _{\mu \nu }}{p_\mu },\;\;\sum\limits_{\mu  = 0}^N {{p_\mu }}  = 1$. By using the identity
\begin{equation}
H = \sum\limits_{\mu \nu } {{p_\mu }H{p_\nu }},
\label{eq:16}
\end{equation}
we calculate the diagonal and off-diagonal matrix elements
(\ref{eq:16}). The term ${p_0}H{p_0}$ corresponds
to the Hamiltonian component acting in the Hubbard
band at the root vector ${\alpha _0} = (0,\sigma )$, etc. It is easy to show that the
equality
\begin{equation}
\sum\limits_\mu  {{p_\mu }} {H_0}{p_\mu } = {H_0}
\label{eq:17}
\end{equation}
is satisfied and that the diagonal elements ${p_\mu }{H_1}{p_\mu }$
describe the hoppings in the band $\mu $ and the off-diagonal
elements ${p_\mu }{H_1}{p_\nu }$ correspond to the hybridization of
the bands $\mu $ and $\nu $.
We introduce the small parameter $\varepsilon\ll1$ and Hamiltonian $\tilde{H}(\varepsilon)$ as
\begin{eqnarray}
&&\tilde H(\varepsilon ) = {\tilde H_0} + \varepsilon {\tilde H_1} ,\nonumber\\
&&{\tilde H_0} = \sum\limits_\mu  {{p_\mu }H{p_\mu }}, {\tilde H_1} = \sum\limits_{\mu \nu } {{p_\mu }H{p_\nu }}
\label{eq:19}
\end{eqnarray}
with the perturbation - interband part $\tilde{H_1}$. We also perform the standard unitary
transformation
\begin{equation}
{H'{(\varepsilon )}} = {e^{ - i\varepsilon S}}\tilde{H}(\varepsilon ){e^{i\varepsilon S}}
\label{eq:20}
\end{equation}
to eliminate the linear (over $\varepsilon$) contributions
to ${\tilde H_1}$. If the matrix $\hat S$ satisfies the equation
\begin{equation}
{\tilde H_1} + i{\left[ {{{\tilde H}_0},\;S} \right]_ - } = 0.
\label{eq:21}
\end{equation}
The transformed Hamiltonian are given by
\begin{equation}
H'(\varepsilon ) = {\tilde H_0} + {{i{\varepsilon ^2}{{\left[ {{{\tilde H}_1},\;S} \right]}_ - }} \mathord{\left/
 {\vphantom {{i{\varepsilon ^2}{{\left[ {{{\tilde H}_1},\;S} \right]}_ - }} 2}} \right.
 \kern-\nulldelimiterspace} 2} .
 \label{eq:22}
\end{equation}
In order to solve (\ref{eq:21}), we multiply each term by
${p_\mu }$ from the left and by ${p_\nu }$ from the right. As a result, we
obtain
\begin{eqnarray}
{p_\mu }H{p_\nu }(1 - {\delta _{\mu \nu }}) &+& i({p_\mu }H{p_\mu })({p_\mu }S{p_\nu })\nonumber \\
&-&i({p_\mu }S{p_\nu })({p_\nu }H{p_\nu }) = 0.
\label{eq:23}
\end{eqnarray}
This equation coincides in form with the corresponding
equation in the work ~\cite{Chao_etal1977} and differs from it only in the dimension
of matrices. Therefore, our derivation of the system of
operators ${p_\mu }$ in the multiband case is actually a
generalization of the method proposed in the work.\cite{Chao_etal1977}
It follows from (\ref{eq:23}) that the diagonal matrix elements
have the form  ${p_\mu }S{p_\mu } = \eta {p_\mu }$, where $\eta $ is a constant.
In order to solve the equation with respect to the
off-diagonal elements ${p_\mu }S{p_\nu }$, according to~\cite{Chao_etal1977}, we
make the approximation ${p_\mu }H{p_\mu } \to {\varepsilon _\mu }$. As a result, the
solution has the form
\begin{equation}
{p_\mu }S{p_\nu } = {{i{p_\mu }H{p_\nu }} \mathord{\left/
{\vphantom {{i{p_\mu }H{p_\nu }} {{\Delta _{\mu \nu }}}}} \right.
\kern-\nulldelimiterspace} {{\Delta _{\mu \nu }}}},\;\;{\Delta _{\mu \nu }} = {\xi _\mu } - {\xi _\nu },
\label{eq:24}
\end{equation}
where the $\xi_{\mu}$ is an energy of $\mu$-th eigen state of the $H_0$ (\ref{eq:2}). The effective Hamiltonian is represented as
\begin{widetext}
\begin{eqnarray}
H'(\varepsilon  = 1) & = & \sum\limits_\mu  {{p_\mu }H} {p_\mu } + \frac{1}{2}\;\sum\limits_{\nu  \ne \mu } {({p_\mu }H} {p_\nu }S - S{p_\mu }H{p_\nu }) \nonumber\\
 &=& \sum\limits_\mu  {{p_\mu }H} {p_\mu } - \frac{1}{2}\;\sum\limits_{\mu  \ne \nu } {\left\{ {\frac{{{{\left[ {{p_\mu }H{p_\nu },{p_\nu }H{p_\mu }} \right]}_ - }}}{{{\Delta _{\mu \nu }}}} + \;\sum\limits_{\scriptstyle \alpha  \ne \mu  \hfill \atop
  \scriptstyle \alpha  \ne \nu  \hfill} {\left[ {\frac{{({p_\mu }H{p_\nu })({p_\nu }H{p_\alpha })}}{{{\Delta _{\nu \alpha }}}} - \frac{{({p_\alpha }H{p_\mu })({p_\mu }H{p_\nu })}}{{{\Delta _{\alpha \mu }}}}} \right]} } \right\}}.
  \label{eq:25}
\end{eqnarray}
\end{widetext}
The calculation of the terms in Hamiltonian (\ref{eq:25}) for
the singlet and triplet bands leads to different results.
The interband transitions through the gap are described
by the commutator
\begin{equation}
{\left[ {{p_0}H{p_\nu },\;\;{p_\nu }H{p_0}} \right]_ - }.
\label{eq:26}
\end{equation}
For the n-th singlet band at the root vector ${\vec{\alpha} _\nu } = ( \bar{\sigma} ,\;nS)$, commutator (\ref{eq:26})
is determined by the operators
\begin{equation}
\sum\limits_{fgij} {\sum\limits_{\sigma \sigma '} {{{\left[ {X_f^{\sigma 0}X_g^{ - \sigma ,ns},\;X_i^{ns, - \sigma '}X_j^{0\sigma '}} \right]}_ - }} }.
\label{eq:27}
\end{equation}
The exchange contribution to the Heisenberg
Hamiltonian has the form
\begin{equation}
{H_A} = \sum\limits_{ij} {J_A^{}\left( {{{\vec R}_{ij}}} \right)} \left( {{{\vec s}_i}{{\vec s}_j} - {\textstyle{1 \over 4}}{n_i}{n_j}} \right),
\label{eq:28}
\end{equation}
where $\vec s$ and ${n_i}$ are the spin operators for $s = 1/2$ and the
number of particles at the $i$-th site, respectively, and
\begin{eqnarray}
{J_A}\left( {{{\vec R}_{ij}}} \right)& = &\sum\limits_n {J_A^{\left( n \right)}} ({\vec R_{ij}})
= \sum\limits_{n = 1}^{{N_s}} {{{{{\left| {t_{ij}^{0,ns}} \right|}^2}} \mathord{\left/
{\vphantom {{{{\left| {t_{ij}^{0,ns}} \right|}^2}} {{\Delta _{ns}}}}} \right.
\kern-\nulldelimiterspace} {{\Delta _{ns}}}}} , \nonumber \\
\;\;{\Delta _{ns}}& = &{E_{ns}} - 2{\epsilon _1}.
\label{eq:29}
\end{eqnarray}
For the $m$-th triplet band, commutator (\ref{eq:26}) is determined
by the terms
\begin{widetext}
\begin{equation}
\left[ {X_f^{\sigma ;0}\left( {X_g^{\sigma ; - m,2\sigma } + {\textstyle{1 \over {\sqrt 2 }}}X_g^{ - \sigma ;m,0}} \right)\left( {X_i^{m,2\sigma ';\sigma '} + {\textstyle{1 \over {\sqrt 2 }}}X_i^{m,0; - \sigma '}} \right)X_j^{0;\sigma '}} \right].
\label{eq:30}
\end{equation}
 \end{widetext}
The ferromagnetic exchange contribution
to the Heisenberg Hamiltonian takes the form
 \begin{equation}
{H_F} = \sum\limits_{ij} {J_B^{}\left( {{{\vec R}_{ij}}} \right)} \left( {{{\vec s}_i}{{\vec s}_j} + {\textstyle{3 \over 4}}{n_i}{n_j}} \right),
\label{eq:31}
\end{equation}
where $J_B^{}\left( {{{\vec R}_{ij}}} \right) = \sum\limits_m {J_B^{\left( m \right)}\left( {{{\vec R}_{ij}}} \right)}  =  - \sum\limits_{m = 1}^{{N_T}} {{{{{\left| {t_{ij}^{0,m}} \right|}^2}} \mathord{\left/
 {\vphantom {{{{\left| {t_{ij}^{0,m}} \right|}^2}} {2{\Delta _m}}}} \right.
 \kern-\nulldelimiterspace} {2{\Delta _m}}}}$ and ${\Delta _m} = {E_m} - 2{\epsilon _1}$. By summing up over all singlet and
triplet bands, we find the following expression for the
effective exchange interaction parameter:
 \begin{equation}
J_{ij} = \sum\limits_{n = 1}^{{N_S}} {{{{{\left| {t_{ij}^{0,ns}} \right|}^2}} \mathord{\left/
 {\vphantom {{{{\left| {t_{ij}^{0,ns}} \right|}^2}} {{\Delta _{ns}}}}} \right.
 \kern-\nulldelimiterspace} {{\Delta _{ns}}}}}  - \sum\limits_{m = 1}^{{N_T}} {{{{{\left| {t_{ij}^{0,m}} \right|}^2}} \mathord{\left/
 {\vphantom {{{{\left| {t_{ij}^{0,m}} \right|}^2}} {2{\Delta _m}}}} \right.
 \kern-\nulldelimiterspace} {2{\Delta _m}}}}.
 \label{eq:32}
 \end{equation}
The origin of the antiferromagnetic contribution resulting from the lowest Zhang-Rice and all excited singlet states is the same as in Hubbard model, it is the superexchange. High energy excited states gives less contribution to the total exchange parameter due to the energy denominator. Nevertheless the number of excited singlets in our five orbital approach $N_S=15$ and triplet $N_T=10$.

\section{\label{sec:IV}Compression dependence of the superexchange interaction\\}

A penultimate line in Tab.\ref{tab:1} shows the superexchange constant $J(P)$ calculated by the formula (\ref{eq:32}) at the five orbital approch. The calculations show that the  $J(P)$ increases by $\sim$ 20\% under hydrostatic 3\%-compression. This result can be
compared with experimental results ~\cite{Aronson_etal1990} obtained for the La214. Under the hydrostatic
100Kbar pressure the $r_{Cu-O}$ reduces by $\sim$ -2\%, while the $J$ increases by $\sim$ 10\%.
There are also studies where the linear dependence of the superexchange on the pressure was observed up to a $P=410$Kbar.\cite{Eremets_etall1991}  According to the dependence of the La214 crystal structure on the pressure we can found the pressure $P\sim 205$Kbar corresponds to the 3\%-deformed material.\cite{Akhar_etal1988} The $J(P)$ at this pressure increases by $\sim$ 18\%.\cite{Eremets_etall1991}
The calculated value $J\thickapprox0.15$eV in the undeformed La214 exceeds the 0.1eV$\div$0.13 ~\cite{Aronson_etal1990,Eremets_etall1991} obtained in experiments
on the two-magnon Raman scattering, but the one agrees well with the $J$ = 0.146eV from the neutron experiments.~\cite{Coldea_etal1990} In contrast under the uniaxial 3\% compression along the c-axis  $J(P_c)$ decreases by -5.7\%, i.e. the superexchange constant changes much weaker.
In both cases, the hydrostatic and anisotropic compression the superexchange constant $J(P)$
correlates with the in-plane hopping paramemters and $dd$-excitation energy $ \delta_s=\varepsilon(^3B_{1})-\varepsilon(A_{1})$ involving the the two-hole states: Zhang-Rice state $A_1$ and
triplet state ${}^3{B_{1}}$ (see Tab.\ref{tab:1}).

\begin{table}
\caption {Single electron energies, hopping parameters, $J(P)$ and $\delta$ for orthorhombic La$214$ (all values except the connecting vectors in $eV$). Here $x^2$, $z^2$, $p_x$, $p_y$, $p_z$ denote Cu-$d_{x^2-y^2}$,
Cu-$d_{3z^2-r^2}$, O-$p_x$, O-$p_y$, O$_{ap}$-$p_z$ orbitals respectively.}
\label{tab:1}
\begin{ruledtabular}
\begin{tabular}{l|c|c|c|c}
 \footnotesize Parameters             &\footnotesize Connecting            &\footnotesize 3\%-compr.               &\footnotesize Undefor-    &\footnotesize 3\%-hydro-\\
\footnotesize                     &\footnotesize  vectors                             &\footnotesize along        &\footnotesize med   &\footnotesize static \\
                                  & \footnotesize                  &\footnotesize c axis  &\footnotesize material &\footnotesize compr. \\
\hline
                                   \footnotesize $\varepsilon_{x^2}$&                        &\footnotesize -2.031     &\footnotesize -1.849   &\footnotesize -1.578\\
                                   \footnotesize $\varepsilon_{x^2}-\varepsilon_{z^2}$&                         &\footnotesize 0.119     &\footnotesize 0.225   &\footnotesize 0.204\\
                                   \footnotesize $\varepsilon_{x^2}-\varepsilon_{p_x}$&	        			          &\footnotesize 0.983    &\footnotesize 0.957    &\footnotesize 1.004\\
                                   \footnotesize $\varepsilon_{x^2}-\varepsilon_{p_y}$&	                        & \footnotesize 0.983    &\footnotesize 0.957    &\footnotesize 1.004\\
                                   \footnotesize $\varepsilon_{x^2}-\varepsilon_{p_z}$&       				          &\footnotesize -0.503    &\footnotesize -0.173    &\footnotesize -0.311\\
\hline
\footnotesize t($x^2$,$x^2$)	 &\footnotesize(-0.493,-0.5)          &\footnotesize-0.173                 &\footnotesize-0.188 &\footnotesize-0.215\\

\hline
\footnotesize t($z^2$,$z^2$)	 &\footnotesize(-0.493,-0.5)           &\footnotesize 0.050                  &\footnotesize 0.054 &\footnotesize 0.062\\

\hline
\footnotesize t($x^2$,$p_x$)	 &\footnotesize(0.246,0.25,-0.02)     &\footnotesize 1.302                 &\footnotesize 1.355 &\footnotesize 1.527\\

\hline
\footnotesize t($z^2$,$p_x$)	 &\footnotesize(0.246,0.25,-0.02)      &\footnotesize-0.547	            &\footnotesize-0.556 &\footnotesize -0.618\\

\hline
\footnotesize t($z^2$,$p_z$)	 &\footnotesize(0,0.04,0.445) 	       &\footnotesize0.851                  &\footnotesize0.773 &\footnotesize 0.875\\

\hline
\footnotesize t($p_x$,$p_y$)	 &\footnotesize(0.493, 0.0) 	       &\footnotesize-0.854                 &\footnotesize-0.858 &\footnotesize-0.935\\
\footnotesize t$'$($p_x$,$p_y$)  &\footnotesize(0,0.5,0.041)          &\footnotesize 0.757                 &\footnotesize 0.793 &\footnotesize 0.862\\

\hline
\footnotesize t($p_x$,$p_z$)	 &\footnotesize(-0.246,-0.21,0.465)    &\footnotesize-0.447                 &\footnotesize-0.391 &\footnotesize-0.423\\
\footnotesize t$'$($p_x$,$p_z$)  &\footnotesize(0.246,0.29,-0.425)    &\footnotesize-0.424	            &\footnotesize-0.377 &\footnotesize-0.408\\
\hline
\footnotesize $J(\Delta J\%)$	 &\footnotesize    &\footnotesize 0.14(-5.7\%)                &\footnotesize 0.15 &\footnotesize 0.18(19.9\%)\\
\hline
\footnotesize $\delta_s$	 &\footnotesize    &\footnotesize 0.82                &\footnotesize 1.33 &\footnotesize 1.45 \\
\end{tabular}
\end{ruledtabular}
\end{table}

As shown on Fig.\ref{fig:6} the antiferromagnetic character of superexchange is maintained even at a hypothetical set of Hamiltonian parameters corresponding the $A_{1}\leftrightarrow{^3B_{1}}$ - singlet-triplet crossover.

\section{\label{sec:VI}Conclusions\\}

To sum up, using LDA+GTB approach we can describe the different compressure dependences $J(P)$ in the undoped La214 and $qp$-spectra on the same footing.

\begin{figure}
\includegraphics{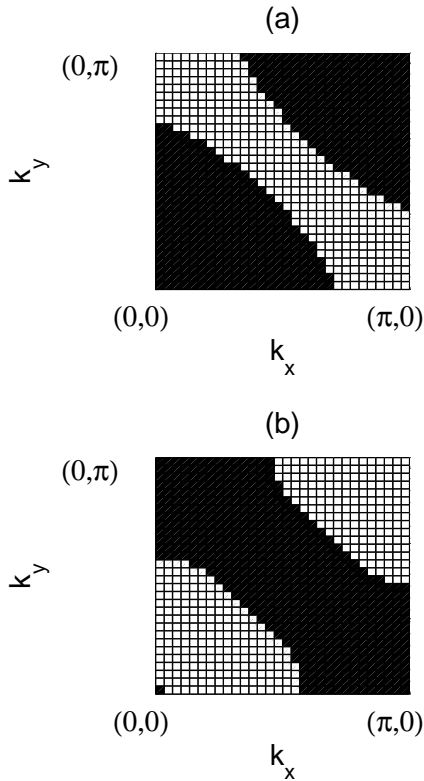}%
\caption{\label{fig:5} (a) and (b): $\vec{k}$ - dependence of the $sign(\delta A_{tot}(P))$ at the $3\%$ - hydrostatic and uniaxial compressions respectively. Black area: (+), white: (-).}
\end{figure}

It is to be stressed that a compressure dynamics of the electron structure and superexchange interaction for the La214 is quite different under the hydrostatic and uniaxial (along c-axial) compressures. As shown on Fig.\ref{fig:5} the $\vec{k}$-depedence of the $sign( \delta A_{tot})$-function  qualitatively reproduces the distribution of $a_1$- and $b_1$-orbital groups over the Brillouin zone. Thus the signs in the photocurrent changes are reversed at the different compressures. Nonetheless
the total photoemission from the $frs$-state: $\int\limits_{\Delta E_{frs}} {\int {{A_{tot}}\left( {\vec k,E,P} \right)\partial E\partial \vec k} } $  integrated over the energy window $\Delta E_{frs}\sim1eV$ slightly decreases by -1.7\% and -2.4\% under the hydrostatic and uniaxial compressures   respectively. Eventually the total photoemission depends on a width of energy window.

Thus even a simple hydrostatic effect on the electronic structure of the anisotropic antiferromagnetic La214 can provide us with useful information about a symmetry of the electronic states. However, currently we do not know anything about the possibilities of a photoemission spectroscopy under a pressure.

A comparison with results for the pressure dynamics of $T_c$ ~\cite{Hardy_etal2010} near optimal hole doping reveals the universal trend of superexchange $J(P)$ together with the $T_c(P)$: ${\raise0.7ex\hbox{${\partial J}$} \!\mathord{\left/
 {\vphantom {{\partial J} {\partial P}}}\right.\kern-\nulldelimiterspace}
\!\lower0.7ex\hbox{${\partial P}$}}>0 $ for isotropic pressure and ${\raise0.7ex\hbox{${\partial J}$} \!\mathord{\left/
 {\vphantom {{\partial J} {\partial P}}}\right.\kern-\nulldelimiterspace}
\!\lower0.7ex\hbox{${\partial P_c}$}}<0 $ for anisotropic one.
The last point is one more argument to the discussion on a nature of the pairing mechanism.  At least at the optimal doping, we conclude that the behavior $J(P)$ is representative of the magnetic pairing interaction in the single $CuO_2$ layer cuprates.\\
\begin{figure}
\includegraphics{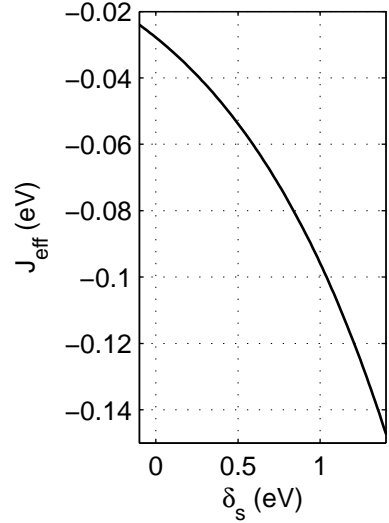}%
\caption{\label{fig:6} A dependence of the superecxhange interaction $J_{eff}$ on the energy of $dd$-excitation $\delta_S$ at the Hamiltonian parameters corresponding to a crossover between the two-hole $A_1$ singlet and $^3B_1$ triplet states.}
\end{figure}
Calculated by the LDA+GTB method the superexchange interaction $J\thickapprox$ 0.15eV in the undeformed La214 agrees well with the $J$ = 0.146eV from neutron experiments.\cite{Coldea_etal1990} The $J(P)$ increases by $\sim$ 20\% under hydrostatic 3\%-compression. In contrast decreasing by $\sim$ -5.7\% the $J$ changes much weaker under uniaxial 3\% compression along the c-axis. In both cases, the hydrostatic and anisotropic compression the superexchange constant $J$ correlates with the in-plane hopping paramemters and the $dd$-excitation energy $ \delta_s$. In fact, at the $\sim$ 205Kbar hydrostatic pressure, the $r_{Cu-O}$  reduces by the -3\%,\cite{Akhar_etal1988} while the $J$ increases by $\sim$ 18\%.\cite{Eremets_etall1991}

Due to the orbital features of the Zhang-Rice state (the same orbitals give the largest contribution to this state and $pd$-hopping) the superexchange $J$ keeps the antiferromagnetic character even at a hypothetical set of Hamiltonian parameters corresponding a crossover of the Zhang-Rice singlet and first excited triplet states (Fig. \ref{fig:6}).

\begin{acknowledgments}
This work was supported by: program of fundamental research of the Russian Academy of Sciences "Quantum mesoscopic and disordered structures" 12-$\Pi$ -2-1002, President of Russia grant ¹ Sh-1044.2012.2 and GK 16.740.12.0731, Presidium of Russian Academy of Science Program 2.16, Siberian and Ural Branch of Russian Academy of Science Projects ¹44 and ¹97, RFFI grants: 11-02-00147, RFFI 12-02-31331, 13-02-01395, 13-02-00358 and Siberian Federal University grant F11.
\end{acknowledgments}

\bibliography{references_gavrichkov}

\end{document}